\documentclass[aps,prl,groupedaddress,showpacs,a4paper]{revtex4}

\usepackage{graphicx}                 
\usepackage{amsmath,amssymb,amsfonts} 

\DeclareFontFamily{OML}{eur}{\skewchar\font127}
\DeclareFontShape{OML}{eur}{m}{n}{<5> <6> <7> <8> <9> gen * eurm <10> <10.95>
  <12> <14.4> <17.28> <20.74> <24.88> eurm10}{}
\DeclareSymbolFont{greek}{OML}{eur}{m}{n}
\DeclareMathSymbol{\micro}{\mathord}{greek}{"16}

\IfFileExists{Fig1.eps}{}{\renewcommand\includegraphics[2][]{\vspace*{5cm}}}

\begin{document}

\title{Space-time nonlinear compression and three-dimensional complex trapping

in normal dispersion}

\author{%
G. Valiulis1,$^1$ J. Kilius,$^{1,2}$ O. Jedrkiewicz,$^2$
A.Bramati,$^2$ S. Minardi,$^2$ C. Conti,$^3$ S. Trillo,$^{3,4}$ A.
Piskarskas,$^1$ and  P. Di Trapani$^2$}

\affiliation{$^1$  Department of Quantum Electronics, Vilnius
University, Sauletekio al.~9,  LT-2040 Vilnius, Lithuania}

\affiliation{$^2$  Istituto Nazionale di Fisica della Materia
(INFM) and Department of Physics,  University of Insubria, Via
Valleggio~11, IT-22100 Como, Italy}

\affiliation{$^3$  Istituto Nazionale di Fisica della Materia
(INFM)-RM3, Via della Vasca Navale 84, 00146 Roma, Italy}

\affiliation{$^4$ Department of Engineering,  Universita' di
Ferrara, Via Saragat 1, 44100 Ferrara, Italy}


\begin{abstract}
  In positive phase-mismatched SHG and normal dispersion,
  a gaussian spatio-temporal pulse transforms spontaneously into a X-pulse,
  underlies spatio-temporal compression and eventually leads to stationary 3-D propagation.
  Experimental and numerical data are provided.
\end{abstract}


 \maketitle

X-pulses (the radial analogous of tilted pulses) are known from
linear optics since they sustain propagation with constant
spatio-temporal profile in normal dispersive media
\cite{Sonajalg}. Here we show that X pulses are formed
spontaneously from a gaussian beam/pulse under SHG with positive
mismatch ($2k_{\omega}>k_{2\omega}$) and normal dispersion.

\begin{figure}[hbt]
  \includegraphics[width=15cm]{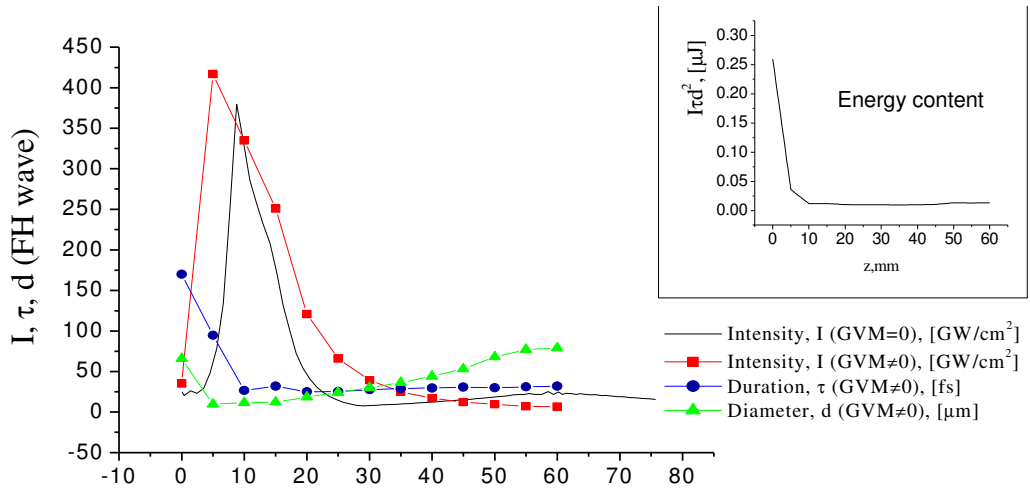}
  \caption{%
    Nonlinear dynamics. Numerical results
  }
\end{figure}
This transformation leads to the relevant dynamics presented by
the numerical results in Figure 1. The achieved spatio-temporal
compression is a genuine effect of the space-time interplay. In
fact, self-focusing and time broadening are expected for
independent spatial and temporal dynamics. During compression the
energy radiates out of the beam center, as shown by the drop in
the energy content ($\rm{E}=\tau d^{2} I$, where $\tau$ and d are
FWHM). After such transient the pulse keeps an almost constant
25\,fs duration (GVM splitting lenght=0.7mm, GVD dispersion
length=6mm, for a gaussian profile). Spatially, a very small
residual diffraction takes place, which let the state to relax
slowly toward the linear regime. Note here the constancy of the
energy content (right), which proves that no radiation occurs
neither in time (due to GVM) nor in space (due to off-axis
components). We verified that a small perturbation of the beam
shape introduces radiation losses. If we switch off the
nonlinearity, than relevant beam diffraction and GVM pulse
splitting occurs. Outside the crystal, the pulse broadens due to
effective anomalous GVD. The scenario indicates the occurrence of
a novel process, which we might call "nonlinear diffusion", that
keeps, at any finite intensity, exact balance among the
dispersions (at all orders) introduced by the material, the
self-phase-modulation and the angular dispersion. Asymptotic
analysis is in progress to see if the field eventually relax to
the exact linear spatio-temporal eigenmode of the double-frequency
field.

\begin{figure}[hbt]
  \includegraphics[width=15cm]{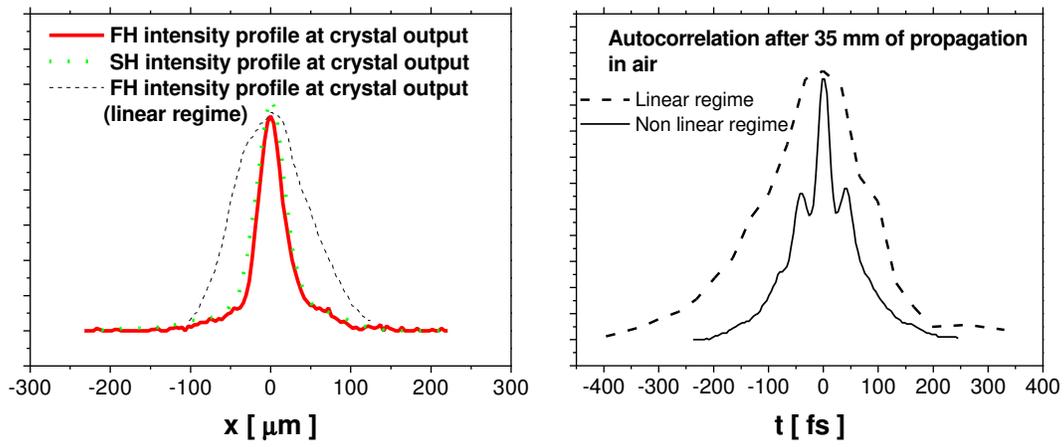}
  \caption{%
    Measured spatial profiles (left) and autocorrelations (right)
  }
\end{figure}
A preliminary experiment done in the same conditions verified the
validity of the model. Figure 2 shows the measured FH and SH
spatial profiles at the output of a 22mm LBO crystal, and the
autocorrelation after 35 mm of further propagation in air. The
agreement with the calculated profiles (see Fig. 3), without free
parameters, is very satisfactory.
\begin{figure}[hbt]
  \includegraphics[width=15cm]{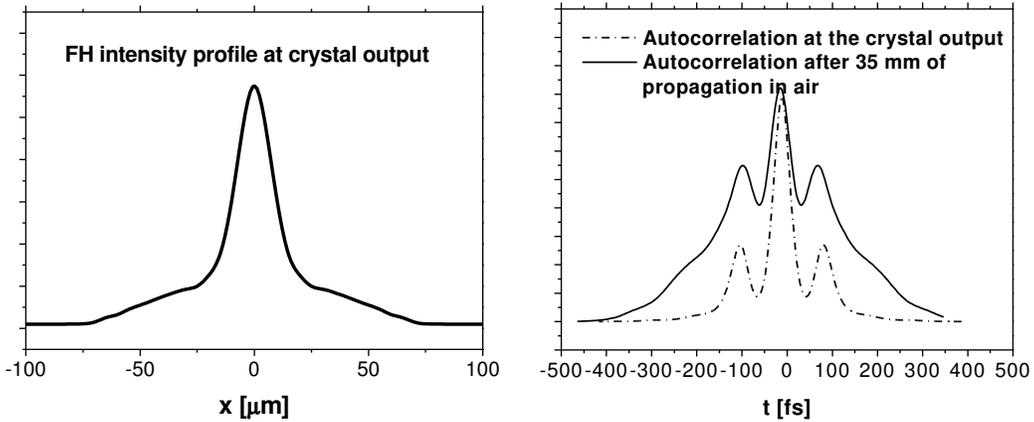}
  \caption{%
    Calculated spatial profile (left) and autocorrelations (right)
  }
\end{figure}

Preliminary calculations indicate that, in the absence of GVM, a
periodical behavior occurs, consistently with the existence of a
X-type pulse with finite envelope (e. g. with finite energy),
which propagate without any relaxation. The intensity evolution of
such a state is plotted as continuos line (no symbols) in Fig. 1.
\cite{Note}.


\begin{thebibliography}{99}
\bibitem{Note} Note: after publication of this paper, the Authors
have discovered that the apparent occurrence of non-linear X-waves
with finite energy in the mentioned, GVM-free, regime was an
artefact related to improper use of reflecting-boundary conditions
in the numerical calculations, owing to the presence of very weak
but fairly extending wave-packet tails
\end{thebibliography}
\end{document}